\documentstyle[aps]{revtex}
\begin{document}
\draft
\title{PLANE  -  CHAIN  COUPLING  IN  YBa$_{2}$Cu$_{3}$O$_{7}$ :
IMPURITY  EFFECT ON  THE  CRITICAL  TEMPERATURE}
\author{R.\ Combescot and X.\ Leyronas}
\address{Laboratoire de Physique Statistique,
 Ecole Normale Sup\'erieure*,
24 rue Lhomond, 75231 Paris Cedex 05, France}
\date{Received \today}
\maketitle

\begin{abstract}
We have studied the effect of impurities on the critical temperature  
for a model of YBCO involving pairing both in the CuO$_{2}$ 
planes and in the CuO chains. In this model pairing in the planes  
is due to phonons, while Coulomb repulsion induces in the chains  
an order parameter with opposite sign. Due to the anticrossing  
produced by hybridization between planes and chains, the order  
parameter changes sign on a single sheet of the Fermi surface  
resulting in nodes in the gap. We find that, in our model, the  
critical temperature is much less sensitive to impurities than  
in standard d-wave models. One reason is that impurities produce  
essentially plane-plane and chain-chain scattering, which does  
not affect the critical temperature. $T_{c } $ is reduced by the  
scattering between parts of the Fermi surface which have opposite  
signs for the order parameter, just as in standard d-wave. In  
our model this is due to plane-chain scattering. We have found  
that this scattering, whatever its origin, will be smaller by  
a factor of order  t $/ E_{F } $ ( that is hybridization coupling  
over Fermi energy ) compared to plane-plane and chain-chain  
scattering. Accordingly the sensitivity of $T_{c } $  to impurities  
in our model is reduced by a similar factor, compared to the  
d-wave situation. In the specific case which we have studied  
in details and which reduces to the two-band model, we have  
found a further reduction of the sensitivity of $T_{c } $  to impurities  
with a behaviour which can vary continuously from s-wave like  
to d-wave like depending on the parameters. We expect a similar  
behaviour and reduction to occur in the general case. 
\end{abstract}

\section {INTRODUCTION} 
The mechanism responsible for high $T_{c } $ superconductivity  
is still highly controversial \cite{1}. While a large part of the  
theoretical effort is based on the hypothesis that Cooper pairs  
are formed in high $T_{c } $ compounds, there is no agreement on the  
physical origin of the pairing interaction nor on the symmetry  
of the pair wavefunction. These two questions are actually intimately  
related. Indeed repulsive interactions, such as produced by  
spin fluctuations \cite{2}, require an order parameter which changes  
sign on the Fermi surface in order to produce pairs. This leads,  
in the simplest hypothesis, to pairing with d-wave symmetry  
if we assume singlet pairing. On the other hand purely attractive  
interactions lead to pairing with s-wave symmetry since any  
change of sign is unfavorable in this case. Therefore an experimental  
determination of the order parameter symmetry should greatly  
help to identify the physical interaction responsible for pair  
formation: although it would not be enough to provide a unique  
identification, it would strongly narrow the remaining possibilities.  
For this reason a large part of the recent experimental work  
has been aimed toward providing a clear signature for the symmetry  
of the order parameter.  \par 
  \bigskip 
Quite surprisingly recent experiments on $YBa_{2}Cu_{3}O_{7} $  
designed with this purpose of identifying the symmetry have  
given clear cut, but contradictory answers \cite{1}. Indeed the observation  
of a sizeable Josephson current \cite{3} in a c-axis tunnelling junction  
between $YBa_{2}Cu_{3}O_{7} $ and  Pb is quite difficult to  
reconcile with a pure d-wave symmetry while it is in full agreement  
with s-wave symmetry. Similarly the fact that there is no angular  
dependence in the critical current of YBCO -  YBCO grain boundary  
junctions in the a-b plane \cite{4} goes clearly in the direction  
of an s-wave interpretation. On the other hand a number of experiments  
are in favor of a d-wave symmetry. Many experiments, including  
tunnelling, NMR, Raman scattering, photoemission, penetration  
depth \cite{1}, have shown the existence of low energy excited states.  
Actually these experiments are compatible with a strongly anisotropic  
s-wave order parameter. Or more simply one may look for extrinsic  
effects and wonder if these states do not arise from surface  
effects or defects, which would provide quasi-normal regions.  
However the existence of a linear T dependence of the penetration  
depth over a large range of temperature, simultaneously in cristals  
and films, with the same slope \cite{5}, makes an extrinsic interpretation  
for all these experiments unlikely ( while this kind of explanation  
may very well be valid for some of them ). Moreover some experiments  
specifically designed to check if the order parameter changes  
sign over the Fermi surface have given positive answers. These  
are the corner SQUID experiments \cite{6} which give a clear indication  
for a change of sign of the order parameter between the a and  
the b axis, and the observation of a spontaneous magnetization  
corresponding to a half magnetic flux quantum in 3 grain-boundary  
Josephson junctions \cite{7} which implies a $\pi  $ shift, in clear agreement  
with d-wave symmetry.   \par 
  \bigskip 
On the other hand the d-wave interpretation is not free of problems.  
For example recent experiments show that some thermodynamical  
superconducting properties are markedly anisotropic ( the a  
and b-axis results are different ). Indeed the penetration depth  
in good YBCO cristals display a strong anisotropy of the penetration  
depth \cite{8}, the specific heat anomaly at $T_{c } $  of the parent  
compound $LuBa_{2}Cu_{3}O_{7} $   ( with same $T_{c } $  as YBCO)  
has a marked anisotropy as a function of the orientation of  
an applied magnetic field \cite{9}. It is difficult to ascribe these  
anisotropies to the weak orthorhombic distorsion, and it is  
more likely that the CuO chains play a significant role in the  
superconducting properties. One of the most conspicuous problem  
of the d-wave interpretation is the weak sensitivity of the  
critical temperature of $YBa_{2}Cu_{3}O_{7} $  to the presence  
of impurities. Indeed any kind of impurities, whether magnetic  
or not, produces in d-wave superconductors \cite{10,11,12} an effect  
analogous to pair-breaking by magnetic impurities in standard  
s-wave superconductors \cite{13}. In particular the critical temperature  
\cite{10,11} decreases rapidly with increasing impurity concentration  
following the Abrikosov - Gorkov law \cite{13}, and superconductivity  
disappears at a critical concentration. In contrast all samples  
of $YBa_{2}Cu_{3}O_{7} $    seem to have a $T_{c } $  around 90  
K. It is difficult to believe that all samples ( including the  
earlier ones ) are clean enough to affect only weakly the critical  
temperature, whereas all microscopic studies show that there  
are always more structural defects than what is generally admitted.  
Moreover ion \cite{14} or electron \cite{15} irradiation experiments have  
shown a rather weak sensitivity of the critical temperature  
of $YBa_{2}Cu_{3}O_{7} $  on the inverse lifetime deduced from  
resistivity measurements \cite{11}. Actually Zn impurities are known  
\cite{16,17} to have a depressing effect on $T_{c }, $ but this can  
be interpreted as a standard pair-breaking effect since the  
environment of a Zn impurity, located in the $CuO_{2} $ planes,  
is known to be magnetic \cite{18}.  \par 
  \bigskip 
In order to solve these contradictions we have proposed recently  
\cite{19} for $YBa_{2}Cu_{3}O_{7} $  a model which mixes s-wave and  
d-wave features. In our model , in addition to the $CuO_{2} $  
planes, the CuO chains play an essential role. The pairing interaction  
within the planes is attractive ( it can be for example produced  
by phonons ). On the other hand the pairing interaction between  
planes and chains is repulsive ( it can be produced by Coulomb  
interaction ). In this way the order parameter has opposite  
signs on the planes and on the chains. Moreover we include the  
hybridization between planes and chains, which corresponds physically  
to take into account the possibility for an electron to jump  
from planes to chains or vice-versa. Naturally the coupling  
responsible for this hybridization is fairly small, but it is  
a well known feature of all band structure calculations \cite{20}.  
It is of importance only when the plane and the chain band intersect.  
In this case it leads to an anticrossing in the dispersion relations,  
and similarly to an anticrossing of the Fermi surfaces, wherever  
the (uncoupled) pieces of the Fermi surface related to plane  
and chain cross. As a result, when we move on a given sheet  
of the Fermi surface, we go from a part which corresponds physically  
to a plane electron, to a part which corresponds physically  
to a chain electron. Since the order parameter 
has opposite signs for plane and chain electrons, this implies  
that the order parameter changes sign on a given sheet of the  
Fermi surface and therefore has nodes on this sheet by continuity.  
Therefore our model provides an order parameter which is quite  
analogous to a d-wave order parameter and it can in this way  
explain \cite{19} all the experiments in favor of d-wave symmetry.  
 On the other hand it does not have a d-wave symmetry since  
the nodes, which occur at the intersection between plane and  
chain bands, have no reason to satisfy $k_{x } $ = $k_{y } $  , and  
the average of the order parameter has no symmetry reason to  
be zero. Hence there is for example no problem with the nonzero  
Josephson current in $YBa_{2}Cu_{3}O_{7} $  - Pb junctions along  
the c - axis \cite{3}. Moreover since the attractive in-plane interaction  
and the repulsive interaction between plane and chain help each  
other, there is no problem to explain the high value of the  
critical temperature of $YBa_{2}Cu_{3}O_{7} $ \cite{21}. Naturally  
our model is specific of $YBa_{2}Cu_{3}O_{7} $  , but we can  
think that it can be generalized to other compounds where the  
role of the chains can be played by other parts of the structure,  
such as the BiO planes in BSSCO. On the other hand there is  
no possibility of this kind in LSCO and accordingly we do not  
expect experiments to display in this compound the same physical  
features as in YBCO.   \par 
  \bigskip 
Our model has common features with many other models. As a two  
band model it is quite similar to the two band models introduced  
by Suhl, Matthias and Walker and others \cite{22,23,24,25,26} to  
describe superconductivity in transition metals. The possibility  
of an interband repulsive interaction in two band models was  
already introduced by Kondo \cite{24}. More recently a two band S-N  
model has been introduced by Abrikosov and Klemm \cite{27} to account  
for Raman scattering data. The idea of an interband repulsion  
has been put forward by various groups \cite{28,29,30,31} recently  
in the context of high $T_{c } $ superconductivity in order to  
show that experiments displaying a change of sign of the order  
parameter did not necessarily imply a spin fluctuation mechanism.  
Our model is also similar to the one proposed by Abrikosov \cite{32},  
where there are attractive and repulsive interactions within  
a single band, leading to a change of sign of the order parameter  
within this band. Finally, as noted above, chain-plane hybridization  
is a standard feature of band structure calculations \cite{20} and  
there have been various suggestions of the importance of the  
chains in the physics of $YBa_{2}Cu_{3}O_{7}, $  whether for  
intrinsic or for extrinsic reasons \cite{33,34,35}, these models  
allowing for the electron to jump between planes and chains.  
  \par 
  \bigskip 
In this paper we consider the effect of impurities on the critical  
temperature in our model. We will actually restrict ourselves  
to non magnetic impurities. Indeed magnetic impurities are easily  
included, but they will naturally lead to pair-breaking and  
produce a $T_{c } $ following the AG law as in other models. Hence  
magnetic impurities can not discriminate between various models  
and we ignore them for simplicity. The conclusion of our study  
is that, in our model, for generic parameters, the critical  
temperature is much less sensitive to impurities than in standard  
d-wave models. Therefore we might say that, with respect to  
the sensitivity of the critical temperature to non magnetic  
impurities, our model behaves as a ''weak''  d-wave model. This  
happens for two independent reasons. First the reduction of  
$T_{c } $ is due to plane-chain scattering, which is weak compared  
to plane-plane scattering. Next the fact that we have a two  
band model provides further possibilities for a weak impurity  
sensitivity. In section II we present our model and we calculate  
the critical temperature of the clean superconductor as a first  
step toward the calculation of the impurity dependence, which  
is dealt with in section III. Finally section IV is devoted  
to a comparison with experimental results and to our conclusion. 

\section{THE  MODEL} 
Let us first define specifically our model. We will describe  
the motion of an electron in planes and chains by the following  
Hamiltonian: 

\begin{eqnarray}
{H}_{0}=\sum\nolimits\limits_{k}^{} {\varepsilon
}_{k}{c}_{k}^{+}{c}_{k}+\sum\nolimits\limits_{k}^{} {\varepsilon
'}_{k}{d}_{k}^{+}{d}_{k}+\sum\nolimits\limits_{k}^{}
{t}_{k}{c}_{k}^{+}{d}_{k}+h.c.
\label{eq1}
\end{eqnarray}
where $c_{k }^{+} $ and $d_{k }^{+} $ are creation operators in  
the plane and in the chain band respectively. The first term  
corresponds to an isolated plane with dispersion relation
$\epsilon _{k }, $  the second one to isolated chains with 
dispersion relation $\epsilon ^{\prime}_{k } $   
and the last term describes hopping between planes and chains.  
Actually this independent electron Hamiltonian does not correspond  
precisely to the situation found in $YBa_{2}Cu_{3}O_{7} $ .  
Indeed $YBa_{2}Cu_{3}O_{7} $ is build by stacking up sets made  
of two $CuO_{2} $  planes and one CuO chains plane. Therefore  
a better description is obtained by the Hamiltonian:  

\begin{eqnarray}
\matrix{{H}_{0}=\sum\nolimits\limits_{n}^{} {\varepsilon
}^{}{c}_{1,n}^{+}{c}_{1,n}+\sum\nolimits\limits_{n}^{} {\varepsilon
}^{}{c}_{2,n}^{+}{c}_{2,n}+\sum\nolimits\limits_{n}^{}
{t}_{p}({c}_{1,n}^{+}{c}_{2,n}+h.c.)\cr \cr +\sum\nolimits\limits_{n}^{}
{\varepsilon '}^{}{d}_{n}^{+}{d}_{n}+\sum\nolimits\limits_{n}^{}
{t}_{c}({c}_{1,n}^{+}{d}_{n}+h.c.)+\sum\nolimits\limits_{n}^{}
{t}_{c}({c}_{2,n}^{+}{d}_{n+1}+h.c.)\cr}
\label{eq2}
\end{eqnarray}
where all quantities are understood to depend on $k_{x } $  and  
$k_{y } $  and summations run also over $k_{x ,y }. $ The indices  
1 and 2 number the $CuO_{2} $  planes and the index  n  numbers  
the stacks. Introducing the even and odd plane band operators  
$c_{\pm } $  by  $c_{1,2} $ = ( $c_{+} $ $\pm  $ $c_{-} $ $)/\surd 2 $ ,
and taking the Fourier transform in the  z  direction, we obtain:  

\begin{eqnarray}
\matrix{{H}_{0}=\sum\nolimits\limits_{k}^{} ({\varepsilon
}^{}+{t}_{p}){c}_{+,k}^{+}{c}_{+,k}+\sum\nolimits\limits_{k}^{} ({\varepsilon
}^{}-{t}_{p}){c}_{-,k}^{+}{c}_{-,k}+\cr \cr +\sum\nolimits\limits_{k}^{}
{\varepsilon '}^{}{d}_{k}^{+}{d}_{k}+\sum\nolimits\limits_{k}^{}
{t}_{+}({c}_{+,k}^{+}{d}_{k}+h.c.)+\sum\nolimits\limits_{k}^{}
{t}_{-}({c}_{-,k}^{+}{d}_{k}+h.c.)\cr}
\label{eq3}
\end{eqnarray}
where a dependence and summation on  $k_{z} $  is now understood,  
and  t $_{+} $ = t $_{c } \surd 2  \cos ( k_{z}  c / 2 )$  and
t $_{-} $  = t $_{c } \surd 2  \sin ( k_{z}  c / 2 )$ , 
with  c  being the size   
of the unit cell along the  z  direction. Band structure calculations  
\cite{20} give a crossing of the chain band Fermi surface with the  
odd plane band Fermi surface, whereas there is no crossing with  
the even plane band. Therefore the even plane band will not  
play an interesting physical role in our model and we will forget  
it for simplicity, but it should clearly be retained in a very  
realistic calculation. This leads us back to the Hamiltonian  
Eq.(1) where the plane band is actually the odd plane band.  
The hopping term has a $k_{z} $ dependence as given by t $_{-} $.
However, if the even plane band happens to cross the chain band, we would have
no problem in extending our model. This   would be particularly easy in the two
extreme cases where the   even and odd plane bands are nearly degenerate 
( t $_{p } $ $<< $   t $_{c } $  ) or well separated 
( t $_{p } $ $>> $ t $_{c } $  )  since we would be led again to the
Hamiltonian Eq.(1).  \par 
  \bigskip 
With respect to pairing interactions, as we have already indicated,  
we take an attractive pairing interaction in the plane which  
can be for example due to phonon exchange. On the other hand  
we assume a repulsive pairing interaction between plane and  
chain. A natural physical origin for this term is the Coulomb  
repulsion, since screening is certainly not very efficient in  
YBCO because it is a weak metal. Moreover screening is probably  
very ineffective along the  z  direction since hopping in this  
direction is small. We note however that hopping is physically  
necessary in order to have a pairing term between planes and  
chains. Finally there is little experimental evidence for an  
attractive pairing interaction in the chains. However we will  
assume its existence for the sake of generality. We include  
also the standard intraband Coulomb repulsion.    \par 
  \bigskip 
Let us first consider the critical temperature of the clean  
superconductor. In a first step we ignore the hopping term   
t $_{k }. $ The Eliashberg equations at the critical temperature  
are :  

\begin{eqnarray}
{\Delta }_{n}{Z}_{n}=\pi T\sum\nolimits\limits_{n}^{} {\lambda }_{n-m}{{\Delta
}_{m} \over \left|{{\omega }_{m}}\right|}-\pi T\mu
\sum\nolimits\limits_{n}^{{\omega }_{c}} {{\Delta }_{m} \over \left|{{\omega
}_{m}}\right|}-\pi Tk\sum\nolimits\limits_{n}^{{\omega }_{c}} {{\Delta '}_{m}
\over \left|{{\omega }_{m}}\right|}
\label{eq4}
\end{eqnarray}

\begin{eqnarray}
{\Delta '}_{n}{Z'}_{n}=\pi T\sum\nolimits\limits_{n}^{} {\lambda
'}_{n-m}{{\Delta '}_{m} \over \left|{{\omega }_{m}}\right|}-\pi T\mu
'\sum\nolimits\limits_{n}^{{\omega }_{c}} {{\Delta '}_{m} \over \left|{{\omega
}_{m}}\right|}-\pi Tk'\sum\nolimits\limits_{n}^{{\omega }_{c}} {{\Delta }_{m}
\over \left|{{\omega }_{m}}\right|}
\label{eq5}
\end{eqnarray}
Here $\Delta _{n } $  and $Z_{n } $  are the order 
parameter and the renormalization   
function at the Matsubara frequency $\omega _{n }. $ Primed quantities  
refer to the chains while unprimed ones correspond to the planes;  
$\lambda _{n } $  and $\lambda ^{\prime}_{n } $  are 
the effective frequency dependent  
interactions due to phonon exchange while $\mu , $ $\mu ^{\prime} $ and k, k'  
are the intraband and interband Coulomb repulsion with cut-off  
$\omega _{c } $  of order of the Fermi energy. While the interactions  
that we consider here have certainly a wavevector dependence  
in the real superconductor, we have no precise idea of what  
they are nor do we have any physical reason to believe that  
this dependence is strong. Therefore, for the sake of simplicity  
and clarity, we have taken isotropic interactions within planes  
and chains. We can now proceed to the usual reduction of the  
cut-off from $\omega _{c } $  to a frequency $\omega _{D} $ 
a few times a typical phonon frequency. 
For $\omega _{n } $    $> $ $\omega_{D} $  , the order  
parameters $\Delta _{n } $  and $\Delta ^{\prime}_{n } $ 
become constants $\Delta _{\infty} $ 
and $\Delta ^{\prime}_{\infty} $   while $Z_{n } $  , $Z^{\prime}_{n } $ 
$\approx  $ 1. These constants are obtained from Eq.(4) and (5) as : 

\begin{eqnarray}
(1+\mu r){\Delta }_{\infty }+kr{\Delta '}_{\infty }=-\pi T\mu
\sum\nolimits\limits_{n}^{{\omega }_{D}} {{\Delta }_{m} \over \left|{{\omega
}_{m}}\right|}-\pi Tk\sum\nolimits\limits_{n}^{{\omega }_{D}} {{\Delta '}_{m}
\over \left|{{\omega }_{m}}\right|}
\label{eq6}
\end{eqnarray}

\begin{eqnarray}
(1+\mu 'r){\Delta '}_{\infty }+k'r{\Delta }_{\infty }=-\pi T\mu
'\sum\nolimits\limits_{n}^{{\omega }_{D}} {{\Delta '}_{m} \over \left|{{\omega
}_{m}}\right|}-\pi Tk'\sum\nolimits\limits_{n}^{{\omega }_{D}} {{\Delta }_{m}
\over \left|{{\omega }_{m}}\right|}
\label{eq7}
\end{eqnarray}
where r = ln $(\omega _{c } $  / $\omega _{D}) $. 
When these results are  carried  
into Eq.(4) and (5), one obtains the same equations, except  
that the cut-off is now $\omega _{D} $  and 
$\mu , $ $\mu ^{\prime} $ and k, k' are replaced  
by renormalized Coulomb interactions $\mu ^*, $ $\mu ^{\prime *} $ 
and k*, k'* given by : 

\begin{eqnarray}
\mu ^*={\mu +(\mu \mu '-kk')r \over (1+\mu r)(1+\mu 'r)-kk'{r}^{2}}
\ \ \ \ \ \ \ \  k^*={k \over (1+\mu r)(1+\mu 'r)-kk'{r}^{2}}
\label{eq8}
\end{eqnarray}
and similar expressions for $\mu ^{\prime *} $ and k'* obtained by exchanging  
primed and unprimed quantities. Naturally one recovers the standard  
Coulomb pseudopotential when there is no interband interaction.  
It seems safe to assume  $\mu \mu ^{\prime} $ - kk' $> $ 0 , 
otherwise we would have pairing from 's-wave' Coulomb interaction alone 
(when   pairing from Coulomb interaction is considered, this is usely  
for higher 'angular momenta' ). In this case renormalization  
does not change the sign of the various quantities. We see that,  
in addition to its direct effect, interband interaction decreases  
the effective intraband repulsion, leading to an increase of  
the critical temperature as it should \cite{36}. In the following  
we will restrict ourselves to the weak coupling limit where  
 $\Delta _{n } $  and $\Delta ^{\prime}_{n } $  
can be taken as constants and mass renormalization  
is negligible. Indeed, as it will be clear in the following,  
we do not expect strong coupling effects to modify qualitatively  
our conclusions. In this case $T_{c } $  depends only on the combinations  
$\lambda ^* $ $\equiv  $ $\lambda _{0} $ - $\mu ^* $  
and $\lambda ^{\prime *} $ $\equiv  $ 
$\lambda ^{\prime}_{0} $ - $\mu ^{\prime *}, $ in addition  
to k* and k'*. From now on we will omit the star for all the  
renormalized coupling constants. Then, if we let x = ln ( $\Omega  $  
/ $T_{c } $ ) where $\Omega  $ is a typical phonon frequency ( or rather  
1/4 of a typical phonon frequency \cite{37} ), $T_{c } $  is obtained  
from :  

\begin{eqnarray}
( x^{-1} - \lambda  ) ( x^{-1} - \lambda ^{\prime} ) = k k^{\prime}
\label{eq9}
\end{eqnarray}
which is the standard two band result \cite{22}. 
Although one expects physically  
that Coulomb interaction reduces the isotope effect, this is  
not obvious from Eq.(8) and (9). This has been shown to be true  
by Kondo \cite{24} in some limiting cases. We show in the Appendix  
that this property holds quite generally. \par
  \bigskip 
As a preliminary to the calculation with impurities, let us  
finally consider how the above calculation of the critical temperature  
is modified when we take into account the hybridization t $_{k } $  
 between planes and chain. As indicated above, we take isotropic  
interactions and order parameters in the planes and in the chains.  
Corresponding to the effective interaction Hamiltonian :  

\begin{eqnarray}
{H}_{int}=-g\sum\nolimits\limits_{k,k'}^{}
{c}_{k'}^{+}{c}_{-k'}^{+}{c}_{-k}{c}_{k}+K\sum\nolimits\limits_{k,k'}^{}
{d}_{k'}^{+}{d}_{-k'}^{+}{c}_{-k}{c}_{k}+h.c.-g'\sum\nolimits\limits_{k,k'}^{}
{d}_{k'}^{+}{d}_{-k'}^{+}{d}_{-k}{d}_{k}
\label{eq10}
\end{eqnarray}
the order parameter satisfy :

\begin{eqnarray}
\Delta =-g\sum\nolimits\limits_{k}^{}
<{c}_{-k}{c}_{k}>+K\sum\nolimits\limits_{k}^{} <{d}_{-k}{d}_{k}>
\label{eq11}
\end{eqnarray}

\begin{eqnarray}
\Delta '=K\sum\nolimits\limits_{k}^{}
<{c}_{-k}{c}_{k}>-g'\sum\nolimits\limits_{k}^{} <{d}_{-k}{d}_{k}>
\label{eq12}
\end{eqnarray}
However because of the hybridization, the plane and chain operators  
$c_{k } $ and $d_{k } $ do not correspond to 
the eigenstates of $H_{0}. $   
The unitary transformation :  

\begin{eqnarray}
c_{i }  = a_{i j }  \gamma _{j }
\label{eq13}
\end{eqnarray}
diagonalizes the Hamiltonian into $H_{0} $  = $\Sigma  $ 
$e_{i } $ $\gamma ^{+}_{i } $  $\gamma _{i } $ 
where we  have set for convenience $c_{1} $ $\equiv  $ c  
and $c_{2} $ $\equiv 
$ d . The    energies $e_{1} $ and $e_{2} $  
of the hybridized bands are given  
by   2 $e_{1,2} $ = $\epsilon  $  + $\epsilon ^{\prime} $ 
$\pm  $ [ ( $\epsilon  $ 
 - $\epsilon ^{\prime} $ $)^{2} $ + 4 t $^{2} $  
$]^{1/2} $  and $a_{11} $ = $a_{22} $ = $[(e_{1} $ - 
$\epsilon ^{\prime})/(e_{1} $   
 - $e_{2})]^{1/2} $  $\equiv  $  $\cos\theta  $  ,
$a_{21} $ = - $a_{12} $ =   
$[(e_{1} $ - $\epsilon )/(e_{1} $  - $e_{2})]^{1/2} $  $\equiv  $  $\sin\theta
$,   with  again the  k  dependence understood everywhere. When we carry out
this transformation in the mean field Hamiltonian   :  

\begin{eqnarray}
{H}^{}={H}_{0}+\Delta \sum\nolimits\limits_{k}^{}
{c}_{k}^{+}{c}_{-k}^{+}+\Delta '\sum\nolimits\limits_{k}^{}
{d}_{k}^{+}{d}_{-k}^{+}+h.c.
\label{eq14}
\end{eqnarray}
we will obtain interband pairing terms, such as  $\gamma _{-k ,1} $  
 $\gamma _{k ,2} $  , coupling the hybridized bands. However we will  
consider that  t $_{k } $ is large enough so that these bands are  
well separated. Specifically this means that we assume $\omega _{D} $  
 $<< $ t $_{k } $  ( otherwise we should make a much more careful  
strong coupling treatment, which would probably not bring anything  
new qualitatively). In this case the above pairing terms will  
be negligible because we can not have two electrons (k,1) and  
(-k,2) at the Fermi surface with opposite wavevectors, but belonging  
to different bands ( we will justify more specifically this  
approximation in the next section ). Therefore the transformation  
Eq.(13) gives :  

\begin{eqnarray}
{H}^{}={H}_{0}+\sum\nolimits\limits_{k}^{} {\delta }_{1,k}{\gamma
}_{1,k}^{+}{\gamma }_{1,-k}^{+}+\sum\nolimits\limits_{k}^{} {\delta
}_{2,k}{\gamma }_{2,k}^{+}{\gamma }_{2,-k}^{+}+h.c.
\label{eq15}
\end{eqnarray}
with  $\delta _{1,k } $ = $\Delta  $ $\cos^{2}\theta _{k } $  + 
$\Delta ^{\prime} $ 
$\sin^{2}\theta _{k } $  
and $\delta _{2,k } $ = $\Delta  $ $\sin^{2}\theta _{k } $  + 
$\Delta^{\prime} $ 
$\cos^{2}\theta _{k } $. Naturally this band diagonal expression for the
Hamiltonian   is obvious once the interband pairing terms are neglected, and  
we could have written it immediately. Our essential point is  
that, starting from isotropic interactions in planes and chains,  
we obtain a specific anisotropy for the order parameter $\delta _{1,k } $  
and $\delta _{2,k } $ . In particular we obtain nodes at the Fermi  
surface since we have managed to have $\Delta  $ and $\Delta ^{\prime} $ 
with opposite signs, and $\cos^{2}\theta _{k } $ goes from 0 to 1 when we
move at the Fermi surface of a given band. Therefore we have an order  
parameter which is d-wave like, although we have assumed an  
attractive pairing in the planes.  \par 
  \bigskip 
We have then as usual $< $ $\gamma  $  $_{-i } $  $\gamma  $  $_{i } $
$> $ = - $(\delta _{i }/2e_{i }) $  
$\tanh (\beta e_{i }/2). $ From Eq.(11),(12) and (13) this leads to:  

\begin{eqnarray}
\Delta =\sum\nolimits\limits_{k}^{} (g{\cos}^{2}{\theta }_{k}-K{\sin}^{2}{\theta
}_{k}){{\delta }_{1,k} \over 2{e}_{1,k}}\tanh({\beta {e}_{1,k} \over
2})+(g{\sin}^{2}{\theta }_{k}-K{\cos}^{2}{\theta }_{k}){{\delta }_{2,k} \over
2{e}_{2,k}}\tanh({\beta {e}_{2,k} \over 2})
\label{eq16}
\end{eqnarray}

\begin{eqnarray}
\Delta '=\sum\nolimits\limits_{k}^{} (g'{\sin}^{2}{\theta
}_{k}-K{\cos}^{2}{\theta }_{k}){{\delta }_{1,k} \over 2{e}_{1,k}}\tanh({\beta
{e}_{1,k} \over 2})+(g'{\cos}^{2}{\theta }_{k}-K{\sin}^{2}{\theta }_{k}){{\delta
}_{2,k} \over 2{e}_{2,k}}\tanh({\beta {e}_{2,k} \over 2})
\label{eq17}
\end{eqnarray}
Since the bands are well separated, the  {\bf k}  summations  
will be around the Fermi surface of each band with cut-off 
$\omega _{D} $   
for $e_{i ,k } $  . The integration perpendicular to the Fermi  
surface will give the standard factor  
x = ln ( 1.13 $\beta  $ $\omega _{D} $) and we are left with summations
along the Fermi surface:

\begin{eqnarray}
\Delta =x\sum\nolimits\limits_{k}^{} (g{\cos}^{2}{\theta
}_{k}-K{\sin}^{2}{\theta }_{k}){\delta }_{1,k}\delta
({e}_{1,k})+(g{\sin}^{2}{\theta }_{k}-K{\cos}^{2}{\theta }_{k}){\delta
}_{2,k}\delta ({e}_{2,k})
\label{eq18}
\end{eqnarray}

\begin{eqnarray}
\Delta '=x\sum\nolimits\limits_{k}^{} (g'{\sin}^{2}{\theta
}_{k}-K{\cos}^{2}{\theta }_{k}){\delta }_{1,k}\delta
({e}_{1,k})+(g'{\cos}^{2}{\theta }_{k}-K{\sin}^{2}{\theta }_{k}){\delta
}_{2,k}\delta ({e}_{2,k})
\label{eq19}
\end{eqnarray}
where the order parameter $\delta _{i
,k } $  should not be confused   with the Dirac function $\delta (e) $. 
For $e_{1,k } $  = 0 we have $\cos^{2}\theta  $   
 $= $  $\epsilon ^{\prime}/(\epsilon  $ + $\epsilon ^{\prime}) $ 
and $\cos^{2}\theta  $ 
 $= $  $\epsilon /(\epsilon  $ + $\epsilon ^{\prime}) $ 
for $e_{2,k } $  
 = 0. Moreover for most of the Fermi surface we have 
$\epsilon  $ $<< $ $\epsilon ^{\prime} $   
or $\epsilon ^{\prime} $ $<< $ $\epsilon  $    
leading to $\cos^{2}\theta  $  $= $ $0 $  or 1. This corresponds  
merely to the result one obtains without hybridization. Subtracting  
from Eq.(18) and (19) their counterpart without hybridization,  
we are left with summations over quantities which are essentially  
nonzero only in the anticrossing region. In this small region  
it is convenient to take, at fixed $k_{z}, $ $\epsilon  $ 
and $\epsilon ^{\prime} $ as new  
variables instead of $k_{x } $  and $k_{y }. $ The Jacobian of the  
transformation is  J = $\partial  $ $(\epsilon ,\epsilon ^{\prime})/
\partial  $ $(k_{x } $ $,k_{y } $ ) = $| $  
${\bf v}_{p } $  x ${\bf v}_{c } $  $| $ where ${\bf v}_{p } $  
and ${\bf v}_{c } $  are the Fermi velocities of the plane and chain band  
at the crossing point. We are then led to evaluate :  

\begin{eqnarray}
\int_{}^{}d\varepsilon d\varepsilon '{f(\varepsilon ,\varepsilon ') \over
{(\varepsilon +\varepsilon ')}^{2}}[\delta ({e}_{1})+\delta ({e}_{2})-\delta
(\varepsilon )-\delta (\varepsilon ')] \nonumber
\end{eqnarray}
with  $f(\epsilon ,\epsilon ^{\prime}) $ = $\epsilon ^{2} $  or 
$\epsilon ^{\prime   2} $ or $-\epsilon \epsilon ^{\prime}. $ This is easily  
done by taking $\epsilon /t $  $_{k } $  and 
$\epsilon ^{\prime}/t $  $_{k } $ as new variables  
and extending the boundaries of the integrals to infinity. The  
result is the same for the three integrals, namely - $\pi  $ $|t $   
$_{k } $  $|. $ When this result is carried into Eq.(18) and (19)  
we find naturally the first order correction to the coupling  
constants $\lambda  $ = $N_{p } $ g , $\lambda ^{\prime} $ = 
$N_{c } $ g' , k = $N_{c } $ K and  
k' = $N_{p } $ K , where $N_{p } $ and $N_{c } $ are the total density  
of states of the (uncoupled) plane and chain bands. The hybridization  
changes these coupling constants respectively into  
$\lambda  $ - $N_{t } $  
(g+K), $\lambda ^{\prime} $ - $N_{t } $ (g'+K) , 
k - $N_{t } $ (g+K) and k' - $N_{t } $  
 (g'+K). Here $N_{t } $  is an effective density of states due  
to the hybridization and given by $N_{t } $ = $\bar{t}$ $/(\pi Jc) $  
where $\bar{t} $ is the average of $|t $  $_{k } $  $| $ 
over $k_{z} $ (we  
have taken into account that there are 4 crossings in the Brillouin  
zone ). Naturally since we have  $N_{t } $  $<< $ $N_{p } $ , 
$N_{c } $ in our model these changes are small.

\section{IMPURITY  EFFECT  ON  THE  CRITICAL  TEMPERATURE}
We consider now the effect on the critical temperature of impurities  
located in the planes and in the chains. We will assume that  
the impurity potential scatters electrons within the planes  
or within the chains, but that it does not scatter them from  
plane to chain. Our physical motivation is naturally that planes  
and chains are physically well separated, which corresponds  
to the fact that the hopping term t $_{k } $  is small. Therefore  
the potential for scattering from plane to chain is reduced  
by a factor of order t $_{k } $ / $t_{0} $ compared to the intraplane  
or intrachain scattering ( where $t_{0} $ is a typical hopping  
term within planes or chains ). This physical point is essential  
for the validity of our result. Including this small plane-chain  
scattering does not make any problem, but for sake of clarity  
and simplicity we will not do it explicitly and we will rather  
come back to this point after we have obtained the effect of
intraplane and intrachain scattering. We will also take an isotropic  
scattering within plane and chain, consistently with what we  
have done for the interactions. Here again we will come back  
later on to the effect of a possible anisotropy. This leads  
us to the following impurity potential :  \par 

\begin{eqnarray}
V={U}_{1}\sum\nolimits\limits_{k,k'}^{}
{c}_{1,k}^{+}{c}_{1,k'}+{U}_{2}\sum\nolimits\limits_{k,k'}^{}
{c}_{2,k}^{+}{c}_{2,k'}
\label{eq20}
\end{eqnarray}
We treat this potential within the Born approximation since,  
for the calculation of the critical temperature, making use  
of the T-matrix approximation is merely equivalent to renormalize  
the scattering potential. There is however another effect produced  
by going beyond the Born approximation, which will be considered  
at the end of this section.  \par 
  \bigskip 
Within the Born approximation and after impurity averaging,  
only contributions to the self-energy corresponding to scattering  
twice on the same impurity are retained \cite{38}. Since the potential  
Eq.(20) does not scatter electrons from plane to chain, the  
impurities contribution to the self-energy has only components  
within planes or chains. They are given by \cite{39}:   

\begin{eqnarray}
{\Sigma }_{i}={{n}_{i}U}_{i}^{2}\sum\nolimits\limits_{k}^{} {\tau
}_{3}{G}_{ii}{\tau }_{3}
\label{eq21}
\end{eqnarray}
where $n_{1} $ and $n_{2} $ are the number of impurities per unit  
volume in the planes and in the chains, $\tau _{i } $  are the Pauli  
matrices in Nambu space and G is the temperature Green's function  
in plane-chain representation. The Green's function is related  
to the self-energy by :  

\begin{eqnarray}
G^{-1}   =  G_{0}^{-1}  - \Sigma                             
\label{eq22}
\end{eqnarray}
where $G_{0} $ is the Green's function in the absence of the impurity  
self-energy . Explicitly we set :  

\begin{eqnarray}
{G}_{ij}^{-1}=(i{\tilde{\Omega }}_{i}-{\varepsilon }_{i}{\tau
}_{3}-{\tilde{\Delta }}_{i}{\tau }_{1}){\delta }_{ij}-t{\tau }_{3}{\sigma
}_{x,ij}
\label{eq23}
\end{eqnarray}

\begin{eqnarray}
{G}_{0,ij}^{-1}=(i\omega -{\varepsilon }_{i}{\tau }_{3}-{\Delta }_{i}{\tau
}_{1}){\delta }_{ij}-t{\tau }_{3}{\sigma }_{x,ij}
\label{eq24}
\end{eqnarray}
where $\omega  $  $= $  $(2n+1) $  $\pi  $  T  is the Matsubara frequency  
( we omit systematically for clarity the index  n  in the Matsubara  
frequency and all frequency dependent quantities ), $\sigma _{x } $  
 is the Pauli matrix in plane-chain space and $\Delta _{i } $  is the  
off-diagonal self-energy due to the pairing interaction ( we  
have set $\epsilon _{1} $ $\equiv  $ $\epsilon  $  and $\epsilon _{2} $ $\equiv  $ 
$\epsilon ^{\prime} $ ). In order to invert  
$G^{-1} $  it is convenient to go the hybridized band representation,  
or more precisely to make the transformation which diagonalizes  
$G^{-1} $  when the off-diagonal part  $\tilde{\Delta}_{i}$ of
the   self-energy is zero. For  $G_{0}^{-1} $    the transformation  
matrix is just $a_{i j } $ $\tau _{0} $  
where $a_{i j } $ is 
the transformation  considered in the preceding section. 
In the present case we  
have merely to formally generalize it by replacing $\epsilon _{i } $   
by $\epsilon _{i } $    - i  $\tilde{\Omega}_{i}$ 
for the particule-particule   
part in Nambu space and by 
$\epsilon _{i } $    + i   $\tilde{\Omega}_{i}$    
in the hole-hole part. We keep the same notation for this transformation,  
but the $a_{i j } $ are now complex. The complete transformation  
matrix is now  
$A_{i j } $  = ( $a_{i j } $ + $a^{*}_{i j } $ 
) $\tau _{0} $  
/2 + ( $a_{i j } $ - $a^{*}_{i j } $ ) $\tau _{3} $ /2 . 
Actually we have  
$\Sigma  $ , $\Delta  $  $<< $ t  since we assume that the 
hybridized bands are   
well separated ( $\Sigma  $ or $\Delta  $  are typically of the same order of  
magnitude in the important cases as it will be clear below ).  
Moreover as usual we will be only concerned with Matsubara frequencies  
$\omega  $ at most of order $\Sigma  $ and $\Delta  $ and therefore 
negligible compared   to t . This implies that, when it 
multiplies a quantity of
order   $\Sigma  $ or $\Delta  $  $, $ we can 
take for $a_{i j } $ its zero order  value,  
namely the real value found in the preceding section. This leads  
to the following expression  $g^{-1} $  = $^{t }A $ $G^{-1} $ A   
for the full Green's function in the hybridized representation  
( $^{t }A $ is the transpose of A ) :   

\begin{eqnarray}
{g}_{ij}^{-1}=-{1 \over 2}[({\tilde{e}}_{i}-{\tilde{e}}_{i}^{{}^*}){\tau
}_{0}+({\tilde{e}}_{i}+{\tilde{e}}_{i}^{{}^*}){\tau }_{3}]{\delta
}_{ij}-{a}_{ki}{\tilde{\Delta }}_{k}{a}_{kj}{\tau }_{1}
\label{eq25}
\end{eqnarray}
where :  

\begin{eqnarray}
2{\tilde{e}}_{1,2}={\varepsilon }_{1}+{\varepsilon }_{2}-i({\tilde{\Omega
}}_{1}+{\tilde{\Omega }}_{2})\pm [{({\varepsilon }_{1}-{\varepsilon
}_{2}-i{\tilde{\Omega }}_{1}+i{\tilde{\Omega
}}_{2}{)}^{2}}^{}+4{t}^{2}{]}^{1/2}\approx 2{e}_{1,2}-2i{\tilde{\omega
}}_{1,2}
\label{eq26}
\end{eqnarray}
and we have set :  

\begin{eqnarray}
2{\tilde{\omega }}_{1,2}={\tilde{\Omega }}_{1}+{\tilde{\Omega }}_{2}\pm
{({\tilde{\Omega }}_{1}-{\tilde{\Omega }}_{2})({\varepsilon }_{1}-{\varepsilon
}_{2}) \over [{({\varepsilon }_{1}-{\varepsilon
}_{2}{)}^{2}}^{}+4{t}^{2}{]}^{1/2}}
\label{eq27}
\end{eqnarray}
Now the dominant contributions to the summations of G over the  
wavevector  {\bf k} will come from the vicinity of the Fermi  
surface of the hybridized bands. In this case either   $\tilde{e}_{1} $  
 is large and all the other matrix elements of $g^{-1} $ are  
small ( of order $\Delta  $ typically ), or  $\tilde{e}_{2} $  is large  
and the other elements are small. It is then easy to see that  
when  $\tilde{e}_{2} $ is large, we obtain g by simply inverting  
the (1,1) block of 
$g^{-1}, $ that is have $g_{i j } $ $\approx  $ $[(g^{-1})_{11}]^{-1}
$    $\delta _{i ,1} $ $\delta _{j ,1} $. 
The other terms of $g_{ij } $  are smaller  
by a factor of order $\Delta  $  over energy separation of the hybridized  
bands ( for fixed {\bf k} ), that is a factor of order $\Delta /t $  
 or less . We have a similar result when  $\tilde{e}_{1} $  is  
large. Taken together this means that we can take $g_{i j } $ as  
block diagonal, that is 
$g_{i j } $ = $g_{i i } $ $\delta _{i j } $ 
with :  

\begin{eqnarray}
{g}_{ii}=-{1 \over {\tilde{e}}_{i}{\tilde{e}}_{i}^{*}+{\tilde{\delta
}}_{i}^{2}}\left({\matrix{{\tilde{e}}_{i}^{*}&{\tilde{\delta }}_{i}\cr
{\tilde{\delta }}_{i}&-{\tilde{e}}_{i}\cr}}\right)
\label{eq28}
\end{eqnarray}
where $\tilde{\delta}_{i}  $  = 
$a^{2}_{i j } $ $\tilde{\Delta}_{j}  $ , that  
is explicitly  
$\tilde{\delta}_{1}  $  =  $\tilde{\Delta}_{1}  $ $\cos^{2}\theta  $   
 + $\tilde{\Delta}_{2}  $  $\sin^{2}\theta  $    and  
$\tilde{\delta}_{2}  $  = $\tilde{\Delta}_{1}  $  
$\sin^{2}\theta  $  + $\tilde{\Delta}_{2}  $  $\cos^{2}\theta . $  \par 
  \bigskip 
We can now calculate the impurity self-energy $\Sigma  $ . 
Since $g_{i j } $   
is block diagonal we have 
$G_{i i } $ = $a^{2}_{i j } $  $g_{j j } $   
. When we carry this result into Eq.(21), we can perform the  
integration perpendicularly to the Fermi surface ( that is integrate  
over $e_{1} $  or $e_{2} $  ) since it converges within an energy  
range of order $\Delta . $ We are then left with summations along the  
Fermi surface, just as in the calculation of the critical temperature  
in the preceding section. This leads to :  

\begin{eqnarray}
{\Sigma }_{i}=-\pi {{n}_{i}U}_{i}^{2}\sum\nolimits\limits_{k}^{}
{a}_{ij}^{2}\delta ({e}_{j}){i{\tilde{\omega }}_{j}{\tau }_{0}-{\tilde{\delta
}}_{j}{\tau }_{1} \over \left|{{\tilde{\omega }}_{j}}\right|}
\label{eq29}
\end{eqnarray}
where we have taken into account that  
$\tilde{\delta}_{i}  $  $<< $ $\tilde{\omega}_{i}  $   
since we are only interested in the calculation of the critical  
temperature. We can then proceed as in the preceding section  
: for most of the Fermi surface we have 
$\epsilon _{1} $ $<< $ $\epsilon _{2} $    
 or $\epsilon _{2} $ $<< $ $\epsilon _{1} $    leading to 
$a_{i j }^{2} $  $= $  $0 $   
or 1. This corresponds to the result without hybridization which  
is :   

\begin{eqnarray}
{\Sigma }_{i}^{0}=-{\Gamma }_{i}(i{\tau }_{0}sign\omega -{{\tilde{\Delta
}}_{i} \over \left|{{\tilde{\Omega }}_{i}}\right|}{\tau }_{1})
\label{eq30}
\end{eqnarray}
where $\Gamma _{i } $  = $\pi  $ $n_{i } $ $N_{i } $  
$U_{i }^{2} $ with $N_{1} $  
$\equiv  $ $N_{p } $  and $N_{2} $ $\equiv  $ $N_{c }. $ 
When we calculate $\Sigma _{i } $   
 - $\Sigma ^{0}_{i } $ we are left with summations 
over nonzero quantities  
only in the anticrossing region. We take then  
$(\epsilon _{1} $ - $\epsilon _{2}) $   
/ t  and $(\epsilon _{1} $ + $\epsilon _{2} $ )/ 2t  
as new variables and extend   
the boundaries of the integrals to infinity. The resulting expression  
simplifies under exchanging $\epsilon _{1} $  and $\epsilon _{2} $  . 
We obtain that the diagonal parts of 
$\Sigma _{i } $  and $\Sigma ^{0}_{i } $  are 
equal  
which implies $\tilde{\Omega}_{i}  $  = $\omega  $ + 
$\Gamma _{i } $  sign 
$\omega  $ . On  
the other hand we find from the off-diagonal part :  

\begin{eqnarray}
{\tilde{\Delta }}_{1}={\Delta }_{1}+{\Gamma }_{1}{{\tilde{\Delta }}_{1} \over
\left|{{\tilde{\Omega }}_{1}}\right|}+{\Gamma }_{1}{{N}_{t} \over {N}_{1}}{2
\over \pi }\int_{0}^{1}{du \over (1-{u}^{2}{)}^{3/2}}[{(1+u{)}^{2}{\tilde{\Delta
}}_{1}+(1-{u}^{2}){\tilde{\Delta }}_{2} \over \left|{(1+u{)}^{}{\tilde{\Omega
}}_{1}+(1-{u}^{}){\tilde{\Omega }}_{2}}\right|}+\{u\rightarrow
-u\}-{2{\tilde{\Delta }}_{1} \over \left|{{\tilde{\Omega }}_{1}}\right|}]
\label{eq31}
\end{eqnarray}
where u = $(\epsilon _{1} $ - $\epsilon _{2})/[(\epsilon _{1} $ - 
$\epsilon _{2})^{2} $  + 4  
$t^{2} $  $]^{1/2} $  , and a similar result for  $\tilde{\Delta}_{2}  $  
.   \par 
  \bigskip 
This equation displays explicitly an essential point for our  
result : the correction to the self-energy due to hybridization  
is small, of order $N_{t } $  / $N_{1,2} $  , compared to the self-energy  
in the absence of hybridization. This correction corresponds  
physically to the scattering from plane to chain induced by  
hybridization. However, without hybridization, we are in the  
standard s-wave situation and impurities do not change the critical  
temperature. The only effect on $T_{c } $ will come from the correction  
due to hybridization. Since this correction is small, we expect  
the change of $T_{c } $ to be small with respect to what one would  
get with a simple d-wave order parameter. Although the integration  
in Eq.(31) can be performed analytically in the general case,  
the result is not simple. Therefore, in order to obtain reasonably  
simple calculations, we will continue our quantitative investigation  
only in the particular case where 
$\Gamma _{1} $  = $\Gamma _{2} $  $\equiv  $  $\Gamma , $   
which implies  
$\tilde{\Omega}_{1} $ = $\tilde{\Omega}_{2} $ $\equiv  $   
$\tilde{\Omega}$    . 
Actually this equality is always valid when the
Matsubara   frequency $\omega  $ is large enough compared to 
$\Gamma _{1} $  and $\Gamma _{2} $. This is in particular always valid when
the impurity concentration   is small enough so that  $\Gamma _{1,2} $  $<< $
$T_{c } $ . Therefore the case we consider is quite reasonable.  \par 
  \bigskip 
In this case the equations for  $\tilde{\Delta}_{1} $ and 
$\tilde{\Delta}_{2} $ simplify into :   

\begin{eqnarray}
{\tilde{\Delta }}_{1}={\Delta }_{1}+{\Gamma }^{}{{\tilde{\Delta }}_{1} \over
\left|{{\tilde{\Omega }}^{}}\right|}+{\Gamma }^{}{{N}_{t} \over
{N}_{1}}{{\tilde{\Delta }}_{2}-{\tilde{\Delta }}_{1} \over \left|{{\tilde{\Omega
}}^{}}\right|}
\label{eq32}
\end{eqnarray}

\begin{eqnarray}
{\tilde{\Delta }}_{2}={\Delta }_{2}+{\Gamma }^{}{{\tilde{\Delta }}_{2} \over
\left|{{\tilde{\Omega }}^{}}\right|}+{\Gamma }^{}{{N}_{t} \over
{N}_{2}}{{\tilde{\Delta }}_{1}-{\tilde{\Delta }}_{2} \over \left|{{\tilde{\Omega
}}^{}}\right|}
\label{eq33}
\end{eqnarray}
We have finally to calculate from Eq.(12) and (13) the self-energies  
$\Delta _{1} $ and $\Delta _{2} $ due to the pairing interactions. We have  
already obtained the off-diagonal part of the Green's function,  
and the calculation is essentially the same as for the off-diagonal  
part of the impurity self-energy. One finds :   

\begin{eqnarray}
{\Delta }_{1}={N}_{1}g\pi T\sum\nolimits\limits_{n}^{} [{{\tilde{\Delta }}_{1}
\over {\left|{\tilde{\Omega }}\right|}^{}}+{{N}_{t} \over
{N}_{1}}{{\tilde{\Delta }}_{2}-{\tilde{\Delta }}_{1} \over {\left|{\tilde{\Omega
}}\right|}^{}}]-{N}_{2}K\pi T\sum\nolimits\limits_{n}^{} [{{\tilde{\Delta }}_{2}
\over {\left|{\tilde{\Omega }}\right|}^{}}+{{N}_{t} \over
{N}_{2}}{{\tilde{\Delta }}_{1}-{\tilde{\Delta }}_{2} \over {\left|{\tilde{\Omega
}}\right|}^{}}]
\label{eq34}
\end{eqnarray}

\begin{eqnarray}
{\Delta }_{2}={N}_{2}g'\pi T\sum\nolimits\limits_{n}^{} [{{\tilde{\Delta
}}_{2} \over {\left|{\tilde{\Omega }}\right|}^{}}+{{N}_{t} \over
{N}_{2}}{{\tilde{\Delta }}_{1}-{\tilde{\Delta }}_{2} \over {\left|{\tilde{\Omega
}}\right|}^{}}]-{N}_{1}K\pi T\sum\nolimits\limits_{n}^{} [{{\tilde{\Delta }}_{1}
\over {\left|{\tilde{\Omega }}\right|}^{}}+{{N}_{t} \over
{N}_{1}}{{\tilde{\Delta }}_{2}-{\tilde{\Delta }}_{1} \over {\left|{\tilde{\Omega
}}\right|}^{}}]
\label{eq35}
\end{eqnarray}
However these equations merely mean that we have to calculate  
the critical temperature with the coupling constants modified  
by the hybridization, as we have found at the end of the preceding  
section. With these coupling constants, Eq.(34) and (35) take  
the simple form :  

\begin{eqnarray}
{\Delta }_{1}=\lambda \pi T\sum\nolimits\limits_{n}^{} {{\tilde{\Delta }}_{1}
\over {\left|{\tilde{\Omega }}\right|}^{}}-k\pi T\sum\nolimits\limits_{n}^{}
{{\tilde{\Delta }}_{2} \over {\left|{\tilde{\Omega }}\right|}^{}}
\label{eq36}
\end{eqnarray}

\begin{eqnarray}
{\Delta }_{2}=\lambda '\pi T\sum\nolimits\limits_{n}^{} {{\tilde{\Delta }}_{2}
\over {\left|{\tilde{\Omega }}\right|}^{}}-k'\pi T\sum\nolimits\limits_{n}^{}
{{\tilde{\Delta }}_{1} \over {\left|{\tilde{\Omega }}\right|}^{}}
\label{eq37}
\end{eqnarray}
From Eq.(32) and (33) we have :  

\begin{eqnarray}
{{\tilde{\Delta }}_{1} \over \left|{{\tilde{\Omega }}^{}}\right|}={{\Delta
}_{1} \over \left|{\omega }\right|}+{{\gamma }_{1}({\Delta }_{2}-{\Delta }_{1})
\over \left|{\omega }\right|(\left|{\omega }\right|+{\gamma }_{1}+{\gamma
}_{2})}
\label{eq38}
\end{eqnarray}

\begin{eqnarray}
{{\tilde{\Delta }}_{2} \over \left|{{\tilde{\Omega }}^{}}\right|}={{\Delta
}_{2} \over \left|{\omega }\right|}+{{\gamma }_{2}({\Delta }_{1}-{\Delta }_{2})
\over \left|{\omega }\right|(\left|{\omega }\right|+{\gamma }_{1}+{\gamma
}_{2})}
\label{eq39}
\end{eqnarray}
where we have set  $\gamma _{1} $  = $\Gamma  $ $N_{t } $  / $N_{1} $  
and $\gamma _{2} $  = $\Gamma  $ $N_{t } $  / $N_{2} $  . 
As expected the inverse lifetime $\Gamma  $   
due to scattering within planes and chains has disappeared from  
the equations, and only are left  $\gamma _{1} $  and $\gamma _{2} $  which  
describe physically the scattering from plane to chain due to  
hybridization. When this is carried into Eq.(36) and (37) and  
the summations on Matsubara frequency is carried out, one finds :  

\begin{eqnarray}
{\Delta }_{1}=\lambda [{\Delta }_{1}x+({\Delta }_{2}-{\Delta }_{1}){k \over
k+k'}K(\rho )]-k[{\Delta }_{2}x+({\Delta }_{1}-{\Delta }_{2}){k' \over
k+k'}K(\rho )]
\label{eq40}
\end{eqnarray}

\begin{eqnarray}
{\Delta }_{2}=\lambda '[{\Delta }_{2}x+({\Delta }_{1}-{\Delta }_{2}){k' \over
k+k'}K(\rho )]-k'[{\Delta }_{1}x+({\Delta }_{2}-{\Delta }_{1}){k \over
k+k'}K(\rho )]
\label{eq41}
\end{eqnarray}
where x = ln ( 1.13 $\omega _{D} $ / $T_{c } $  ) and 
$K(\rho ) $ = $\psi (1/2 $ +  $\rho  $ ) - $\psi (1/2) $ , with 
$\rho  $ = $(\gamma _{1} $  + $\gamma _{2})/(2\pi T_{c }), $   
and we have used the fact that in our case k / k' = $\gamma _{1} $  
 / $\gamma _{2} $ . Eq.(40) and (41) are essentially what one obtains  
for the effect of the impurities in a two band model \cite{26}. This  
is easy to understand physically since, because of hybridization,  
the in-plane scattering potential allows effectively an electron  
to scatter from plane to chain for example. However our essential  
point is that the corresponding effective inverse scattering  
time $\gamma  $ $\equiv  $ $\gamma _{1} $  + $\gamma _{2} $    
is strongly reduced, compared  
to plane-plane or chain-chain scattering, because the hybridization  
is small. We note also that, in the general case, our equations  
Eq.(31) do not reduce to a simple two-band model, although we  
expect the physics to be similar.   \par 
  \bigskip 
Let us briefly review the existing literature \cite{26} for the consequences  
of Eq.(40) and (41) for the critical temperature, and add some  
results for the case which is of interest for us, namely the  
general situation with repulsive interactions. From Eq.(40)  
and (41) one gets an equation for  x  which is more conveniently  
rewritten as :  

\begin{eqnarray}
K(\rho )={(x-{x}_{0})(x-{x}_{1}) \over x-{x}_{2}}
\label{eq42}
\end{eqnarray}
Here $x_{0} $  and $x_{1} $  are the solutions of Eq.(9). More precisely  
Eq.(9) has two physically relevant solutions when  
$\lambda \lambda ^{\prime} $ $> $ kk'.  
The lower one $x_{0} $  gives the physical critical temperature.  
The higher one $x_{1} $  corresponds to an unstable superconducting  
state where the order parameter has the same sign in the two  
bands, and therefore the interband coupling decreases the critical  
temperature with respect to the uncoupled situation instead  
of increasing it ( in the uncoupled limit $x_{0} $ = $1/\lambda  $ and  
$x_{1} $ = $1/\lambda ^{\prime} $ ; we assume for example 
$\lambda  $ $> $ $\lambda ^{\prime} $ 
\cite{41}).  When  
$\lambda \lambda ^{\prime} $ $< $ kk', the interband coupling is 
too strong for the state  
with the ill-chosen signs of the order parameter to exist and  
the unstable solution disappears ( there is still a mathematical  
solution with $x_{1} $  $< $ 0 but it must be rejected physically  
because it does not satisfy the weak coupling condition x $>> $  
1 anymore ). Finally we have  
( $\lambda \lambda ^{\prime} $ - kk' ) $x_{2} $  = 
( $\lambda  $  
k + $\lambda ^{\prime} $ k' + 2 kk' )/ (k+k') in Eq.(42). 
It is easy to see that   one has always 
$x_{0} $  $< $ $x_{2} $ $< $ $x_{1} $  for 
$\lambda \lambda ^{\prime} $ $> $ kk', while 
$x_{1} $  $< $ $x_{2} $ $< $ $x_{0} $  is satisfied for 
$\lambda \lambda ^{\prime} $ $< $ kk'  
. Now for fixed $\gamma  $ , $K(\rho ) $ is a positive, increasing function  
of x + ln $(\gamma /\omega _{D}) $ , with 
$K(\rho ) $ $\rightarrow  $ 0 for 
x $\rightarrow  $ - $\infty  $  and $K(\rho ) $  
$\approx  $ x + ln $(\gamma /\omega _{D}) $ for x 
$\rightarrow  $ $\infty  $ 
. Then it is easy to see graphically  
that, starting from $x_{0} $  for $\gamma  $ = 0, x as given by Eq.(42)  
is an increasing function of $\gamma  $ . Therefore, in all possible  
cases, the critical temperature decreases when the impurity  
concentration increases.  \par 
  \bigskip 
For $\lambda \lambda ^{\prime} $ $> $ kk'  
the increase of  x  saturates at  $x_{2} $ leading  
to a large concentration limit $T_{c }^{\infty} $  for the critical  
temperature given, as found by Kusakabe \cite{26}, by :  

\begin{eqnarray}
\ln({1.13{\omega }_{D} \over {T}_{c}^{\infty }})={2kk'+\lambda k+\lambda
'k' \over (\lambda \lambda '-kk')(k+k')}
\label{eq43}
\end{eqnarray}
This saturation corresponds physically to the situation where  
strong impurity scattering has made the order parameter completely  
isotropic $\Delta _{1} $ = $\Delta _{2 } $  while 
( $\Delta _{1} $ - $\Delta _{2} $ )  
$K(\rho ) $ takes an independent nonzero value. The result Eq.(43)  
is then easily rederived directly from Eq.(40) and (41). Therefore  
in this case the superconducting state survives very strong  
impurity scattering by becoming isotropic. This situation is  
analogous to the fate of very dirty anisotropic s-wave superconductors  
\cite{40} and naturally quite different from the standard behaviour  
of a d-wave superconductor. We note incidentally that, when  
one goes from the pure system to the very dirty one, one goes  
from $\Delta _{1} $ $\Delta _{2} $  $< $ 0 to 
$\Delta _{1} $ $\Delta _{2} $  $> $ 0, which implies  
that, if  $\lambda  $  $+ $  k' $> $ $\lambda ^{\prime} $ + k  
for example, the chains become  
gapless $\Delta _{2} $ = $0 $  $   at $ some specific impurity concentration  
( given from Eq.(40) and (41) by 
$K(\rho ) $ = $(k+k')/(\lambda \lambda ^{\prime} $ - kk')  
and x = $(\lambda ^{\prime}+k)/(\lambda \lambda ^{\prime} $ - kk') ). 
However this case $\lambda \lambda ^{\prime} $ $> $ kk'  
requires not too small coupling constants $\lambda  $ and 
$\lambda ^{\prime} $ in both  
bands. In YBCO it is not clear at all that the Cu0 chains alone  
have a superconducting tendency. If we have 
$\lambda ^{\prime} $ $\approx  $ 0, the condition  
$\lambda \lambda ^{\prime} $ $> $ kk' cannot be satisfied.  \par 
  \bigskip 
For $\lambda \lambda ^{\prime} $ $< $ kk'  
there is no saturation in the increase of   
x  and the critical temperature goes to zero when the impurity  
concentration is increased, as found by Kusakabe \cite{26}. However  
when we substitute in Eq.(42) the large  x  behaviour of $K(\rho ), $  
we obtain :  

\begin{eqnarray}
x-{x}_{0}=\ln{{T}_{c}^{0} \over {T}_{c}^{}}={({x}_{0}-{x}_{2})\ell
n(1.13\gamma /{T}_{c}^{0}) \over {x}_{2}-{x}_{1}-\ell n(1.13\gamma
/{T}_{c}^{0})}
\label{eq44}
\end{eqnarray}
where $T_{c }^{0} $  is the critical temperature without impurities.  
Therefore $T_{c } $  goes to zero for a critical impurity concentration  
and we find in this case a behaviour similar to the standard  
d-wave result. The corresponding critical value  $\gamma _{c } $     
for $\gamma  $  is given by :  

\begin{eqnarray}
\ln{1.13{\gamma }_{c} \over {T}_{c}^{0}}={x}_{2}-{x}_{1}={1 \over
2(kk'-\lambda \lambda ')}\left[{\sqrt {(\lambda -\lambda
'{)}^{2}+4kk'}-{(\lambda -\lambda ')(k-k')+4kk' \over k+k'}}\right]
\label{eq45}
\end{eqnarray}
The right hand side of Eq.(45) is positive and it goes from  
zero ( for $\lambda  $ - $\lambda ^{\prime} $ = k - k' , or k' = 0 ) 
to infinity ( for  
 $\lambda  $ $\lambda ^{\prime} $ = k  k' ). 
Therefore the lowest possible value 
of $\gamma _{c } $  
 is the standard d-wave result $\gamma _{c } $  = .88 $T_{c }^{0} $  ,  
but it can easily be much higher since it is given by the exponential  
of the r.h.s. of Eq.(45). From this, one would conclude that,  
for a given $\gamma , $ our model is much less sensitive to impurities  
than standard d-waves. However when one looks at the behaviour  
of $T_{c } $ as a function of $\gamma , $ one sees that this large value  
of $\gamma _{c } $ is obtained at the end of a long tail where the value  
of $T_{c } $  is already quite small, as it can be seen in Fig.1.  
Therefore in order to fully characterize the sensitivity of  
$T_{c } $  to impurities, we have also to look at the initial slope  
found for low impurity concentration.  \par 
  \bigskip 
In this regime we have from Eq.(42) :  

\begin{eqnarray}
1-{{T}_{c} \over {T}_{c}^{0}}={\pi  \over 4}{\gamma  \over
{T}_{c}^{0}}{{x}_{0}-{x}_{2} \over {x}_{0}-{x}_{1}}={\pi  \over 4}{\gamma  \over
{T}_{c}^{0}}[{1 \over 2}+{(\lambda -\lambda ')(k-k')+4kk' \over 2(k+k')\sqrt
{(\lambda -\lambda '{)}^{2}+4kk'}}]
\label{eq46}
\end{eqnarray}
which is the result of Moskalenko and Palistrant \cite{26}. As they  
indicated, ( $x_{0} $ - $x_{2} $ ) / ( $x_{0} $ - $x_{1} $ ) goes from  
0 to 1, which implies that the decrease of $T_{c } $  is always  
less than for the pure d-wave case. The d-wave result is obtained  
in the limiting case   $\lambda  $ - $\lambda ^{\prime} $ = k - k' or 
k' $\rightarrow  $ 0. 
On the other hand for k $\rightarrow  $ 0 ( and fixed k' ), the critical  
temperature does not depend anymore on the impurity concentration  
( this result is actually valid whatever the impurity concentration  
as it can be seen from Eq.(42)) . This limiting case is not  
unphysical since it corresponds to a situation where the density  
of states $N_{c } $  in the chains would be very small. We would  
still have a d-wave type order parameter with change of sign  
and nodes, but the critical temperature would be completely  
controlled by the planes and the chains would simply follow.  
Therefore we have the surprising result of a superconducting  
phase with a d-wave type order parameter which has a critical  
temperature insensitive to impurities just as an s-wave superconductor.  
However it is fair to say that we do not expect such a limiting  
situation to occur in YBCO.  Hence we believe that in our case  
Eq.(46) leads to a sensitivity to impurity which is always somewhat  
reduced with respect to the standard d-wave case, but not by  
a large factor. An example of the general behaviour of $T_{c } $  
 $   / $ $T_{c }^{0} $  as a function of our scattering rate $\gamma  $ is  
given in Fig.1 where we have chosen the parameters 
$\lambda  $ = 1, $\lambda ^{\prime} $   
= 0 , k' = 0.5 and k takes the values 0.01 , 0.1 , 0.2 , 0.5  
, 1.5 and 5  ( k = 1.5 gives exactly an Abrikosov Gorkov law  ).  \par 
  \bigskip 
In order to conclude this section, let us now come back to the  
various terms we have omitted from the beginning. We did not  
take into account direct plane-chain scattering by impurities.  
As we mentionned, we expect physically this scattering to be  
reduced by a factor $N_{t } $  / $N_{p ,c } $  
compared to plane-plane   
or chain-chain scattering which we have treated. Including this  
plane-chain term is just what is done in the standard two-band  
model when the effect of impurities on  $T_{c } $   is calculated.  
Since we have shown that our calculation lead us to the two-band  
model, including direct plane-chain scattering would just give  
us an additional term of the same order as the one we have found  
( it would appear as an additional term in $\Sigma _{0} $  ). Therefore  
our conclusions with respect to the sensitivity of $T_{c } $    
to impurities are unchanged. We have also omitted mixed plane-chain  
terms $\Delta _{12} $  for the order parameter. This is justified  
by our finding that, when the hybridized bands are well separated,  
the dominant contributions of  g  come from the block diagonal  
parts $g_{11} $ and $g_{22} $ . Hence we have for example $G_{12} $  
= $a_{11} $ $g_{11} $ $a_{21} $ + $a_{12} $ $g_{22} $ $a_{22} $ .  
However the products $a_{11} $ $a_{21} $  and $a_{12} $ $a_{22} $  
are significantly different from zero only in the anticrossing  
region. When we calculate the off-diagonal self-energy, we have  
to sum  $G_{12} $  over {\bf k}. Accordingly the result  $\Delta _{12} $  
 will be smaller by a factor of order $N_{t } $  / $N_{p ,c } $   
compared to $\Delta _{1} $ and $\Delta _{2}, $ 
which justifies our approximation.   
Finally we note that going beyond the Born approximation will  
not only renormalize the impurity cross section. It will also  
introduce plane-chain terms $\Sigma _{12} $  in the impurity self-energy  
because, after scattering on a plane impurity, an electron can  
go to the chains through $G_{12} $ and scatter on a chain impurity.  
But we have just seen that, after summation of $G_{12} $  over  
{\bf k} , one obtains a result which is smaller by a factor  
 $N_{t } $  / $N_{p ,c } $  . 
This leads to a $\Sigma _{12} $  which is   
smaller than $\Sigma _{1} $  or $\Sigma _{2} $  by the same factor and hence  
negligible.

\section{DISCUSSION}
The conclusion of our study is that, in our model, the critical  
temperature is much less sensitive to impurities than it is  
in standard d-wave models. One basic reason is that we expect  
physically impurities to produce essentially plane-plane and  
chain-chain scattering. Just as in s-wave superconductors, this  
scattering does not affect the critical temperature. On the  
other hand $T_{c } $ is reduced by scattering between parts of  
the Fermi surface which have opposite signs for the order parameter,  
just as in standard d-wave. In our model this is due to plane-chain  
scattering. We have found that this scattering, whatever its  
origin, will be smaller by a factor of order  t / $E_{F } $ ( that  
is hybridization coupling over Fermi energy ) compared to plane-plane  
and chain-chain scattering. From the band structure calculations  
we expect this factor to be typically somewhere between 0.1  
and 0.3 . Therefore the sensitivity of $T_{c } $  to impurities  
in our model is reduced by a similar factor, compared to the  
d-wave situation. In the specific case which we have studied  
in details and which reduces to the two-band model, we have  
found a further reduction of the sensitivity of $T_{c } $  to impurities  
with a behaviour which can vary continuously from s-wave like  
to d-wave like depending on the parameters. From our discussion  
of its physical origin, we expect a similar behaviour and reduction  
to occur in the general case.  \par 
  \bigskip 
It is unfortunately not possible to make quantitative comparison  
with experiments. Indeed it seems surprisingly quite difficult  
\cite{18} to avoid a magnetic character for substitutional impurities  
in YBCO. Naturally this occurs when the isolated impurity atom  
itself has a magnetic moment. But this happens also when a non  
magnetic impurity acquires a magnetic moment due to its interaction  
with the environment, as it is the case for Zn for example \cite{18}.  
Naturally an impurity with a magnetic character will produce  
pair-breaking leading to an Abrikosov-Gorkov like  law in any  
model, in agreement with what is observed experimentally. Hence  
these kind of experiments can not be used directly to eliminate  
theoretical models. Nevertheless, after these words of caution  
with respect to a simple-minded interpretation of impurities  
experimental results, we note that the reduction of $T_{c } $ down  
to 13 K for 8\% Zn is obtained by an increase of the residual  
resistivity \cite{18} by a factor of order 10. The corresponding  
$\hbar$ / $\tau  $ should be then of order 20 $T_{c } $ ( see below  
). This is much more than what is necessary to destroy superconductivity  
within a d-wave model according to the Abrikosov and Gorkov  
law. Moreover it is useful to plot the results of Ref.17 and  
Ref.18 for the critical temperature as a function of the residual  
resistivity. This is done on Fig.2. It can be seen that the  
behaviour of some of our results found in Fig.1 is quite similar  
to the experimental results ( we assume naturally that the residual  
resistivity is proportional to the scattering rate due to impurities  
). We have naturally enough adjustable parameters to fit them  
nicely. However such a fit would be rather meaningless because,  
in addition to the magnetic problem, the scatter in the data  
is rather important at low impurity content and the interpretation  
of the data for larger concentration is uncertain ( localisation  
effects which are not taken into account in our theory might  
play a significant role).      \par 
  \bigskip 
If we turn to irradiation experiments, whether by electrons  
or by light ions, the interpretation is also not an easy one.  
It is known that high $T_{c } $ compounds have a critical temperature  
much more sensitive to irradiation than standard superconducting  
materials. However, although it is likely that most of the created  
defects do not have a magnetic character, we can not eliminate,  
from our knowledge on substitutional impurities, the possibility  
that some are magnetic. The experimental evidence is controversial  
in this respect \cite{15}. Then there is some evidence that localisation  
effects might be important since a metal - insulator transition  
is observed in YBCO under light ion irradiation, and there is  
no intermediate normal phase between the superconducting phase  
and the insulating one \cite{14}. This sensitivity to localisation  
is easy to understand because of the two-dimensional nature  
of the $CuO_{2} $  planes. Localisation effects are not included  
in our theoretical study. It is also not clear at all that the  
created disorder can be considered as homogeneous \cite{12} since  
the resistivity measurements do not always display a sharp drop  
at the critical temperature. Similarly one might expect that  
the defects created by irradiation are randomly distributed  
at the microscopic scale, but it is actually quite likely that  
the chains are more sensitive to irradiation than the planes.  
There is finally the obvious problem of having an experimental  
determination of the quasiparticle lifetime produced by disorder.  
Measuring the increase in resistivity appears the best way to  
do it \cite{11} although it is far from perfect since, for example,  
it measures at best transport lifetime which we expect to be  
somewhat larger than quasiparticle lifetime.   \par 
  \bigskip 
Notwithstanding the above problems let us try to interpret the  
irradiation experiments with our theoretical model, just to  
see what comes out. From Drude's law, with a typical resistivity  
of 100 $\mu \Omega  $ cm at $T_{c } $ 
and a plasma frequency of 1.1 eV \cite{15},   
we have a typical inverse lifetime 
$\hbar $ / $\tau  $ $\approx  $ 2 $T_{c }. $   
Since superconductivity disappears in d-wave when the inverse  
lifetime due to impurities is of order 2 $T_{c } $  we would expect  
that an increase of resistivity of 100 $\mu \Omega  $ cm leads to the  
suppression of $T_{c } $  \cite{11}. This corresponds roughly to a decrease  
of $T_{c } $  of 1K per $\mu \Omega  $ cm (the AG law is essentially linear  
). The experimental results of Ref.15 give a linear decrease  
of $T_{c } $    with respect to resistivity with a slope 0.3 K  
/ $\mu \Omega  $ cm. In Ref.14 there is an upward curvature at low resistivity  
with a maximal initial slope of 0.1 K/ $\mu \Omega  $ cm.
Since in our model a reduction by a typical factor 1/10 with  
respect to the d-wave result corresponds to a typical choice  
of our parameters, we see that this last experimental result  
agrees with our expectation. But the result of Ref.14 could  
easily be explained for example by a larger value of the hybridization  
energy or by a suitable choice of the other parameters of our  
two-band model. Quite generally we have enough parameters to  
vary in our model so we can get easily agreement with these  
various experimental results. However it must also be kept in  
mind that, as discussed above, there are other possible physical  
processes which we have not taken into account and which will  
add up to produce a faster decrease of the critical temperature  
with the resistivity. A clear example of this is found in Ref.14  
where a more rapid decrease is found for a set of samples and  
attributed to extrinsic effects, while good samples show an  
average decrease of 0.03 K/ $\mu \Omega  $ cm. Therefore we can consider  
an experimental result as an upper bound for our theoretical  
result, but it may quite well be larger than what we find. Clearly  
it is more difficult to explain in a d-wave model the slow dependence  
of $T_{c } $ on resistivity found in Ref.14 than it is in our model  
to explain the somewhat stronger dependence found in Ref.15.  \par 
  \bigskip 
In conclusion we have seen that our model is quite coherent  
with the present experimental evidence. Naturally, with respect  
to this problem of the effect of impurities on the critical  
temperature, it would be much better to have experiments providing  
stronger constraints on theoretical models, but this might prove  
difficult to achieve.   \par 
  \bigskip 
We are extremely grateful to N. Bontemps, P. Monod and J. Lesueur  
for very useful discussions on the experimental situation. \par 
\bigskip
\bigskip

{\bf APPENDIX} \par
\bigskip 
Let us vary Eq.(9) with respect to the phonon frequencies 
$\omega _{D} $  
 and $\Omega  $ . We assume for example 
$\delta r $ = - $\delta \omega _{D} $ 
 / $\omega _{D} $  
 = - $\delta \Omega  $  $   / $ $\Omega  $  $< $  $0. $  
The variation of Eq.(9) gives :   

\begin{eqnarray}
\delta y   ( 2y - \lambda ^*  - \lambda ^{\prime *} ) = 
\delta (  k^*k'^* ) - \delta \mu ^*   ( y - \lambda ^{\prime *}) 
- \delta \mu ^{\prime *}   ( y - \lambda ^*) 
\label{eqA1}
\end{eqnarray}
where $y = x^{-1} $  . The isotope effect is reduced if we show  
that  $\delta y $ $< $ 0 . We set  
$D^{-1} $  = ( $1+\mu r $ ) ( $1+\mu ^{\prime}r $ ) - $kk'r^{2} $  
 . We have $\delta$ ( $ k^*k'^*$ ) =$ 2k^*k'^*$ $\delta D $ / D ,   and 
$\delta \mu ^* $ = $\mu ^* $  
$\delta D $ / D + D ( $\mu \mu ^{\prime} $ - kk' ) $\delta r $ $> $ 
$\mu ^* $ $\delta D $ 
/ D ( we assume $\mu \mu ^{\prime} $  
- kk' $> $ 0 , which implies $\mu ^*\mu ^{\prime *} $ - $k^*k'^*>0 $ ). 
Since we have   
y $> $ $\lambda ^* $ and  y $> $ $\lambda ^{\prime *}, $ and 
$\delta D $ $> $ 0 it is enough to prove that :   

\begin{eqnarray}
2k^*k'^*  - \mu ^*   ( y - \lambda ^{\prime *})  - \mu ^{\prime *} 
( y - \lambda ^*) <  0 
\label{eqA2}
\end{eqnarray}
The left hand side is zero when the variable y is equal to Y  
= ($ 2k^*k'^*$  + $\mu ^* $  $\lambda ^{\prime *} $ + $\mu ^{\prime *} $  
$\lambda ^*)/(\mu ^*+\mu ^{\prime *}). $ 
 We find that  
$y_{1} $ $< $ Y $< $ $y_{0}, $ since we have :  

\begin{eqnarray}
(\mu ^*+\mu ^{\prime *})^{2}  [ ( Y - \lambda ^* )
 ( Y - \lambda ^{\prime *}) - k^*k'^* ] =   \nonumber \\
  - k^*k'^* [\lambda ^*-\lambda ^{\prime *}+\mu ^*-\mu ^{\prime *}]^{2} 
- (\mu ^*\mu ^{\prime *}  - k^*k'^*) 
[ (\lambda ^*-\lambda ^{\prime *})^{2} + 4  k^*k'^* ] <  0 
\label{eqA3}
\end{eqnarray}
where $y_{0} $ =1/ $x_{0} $ and $y_{1} $ =1/ $x_{1} $ . Since we have  
y $> $ $y_{0} $ $> $ Y , Eq.(48) is satisfied and the isotope effect is  
indeed reduced. There is no reduction only in the limiting case  
$\mu \mu ^{\prime} $ = kk' and $\lambda ^*+\mu ^*=
\lambda ^{\prime *}+\mu ^{\prime *}. $ \par
\bigskip
\bigskip

* Laboratoire associ\'e au Centre National
de la Recherche Scientifique et aux Universit\'es Paris 6 et Paris 7. \par

\begin{figure}
\caption{Relative variation of the critical temperature from  
Eq.42 as a function of our reduced effective  relaxation rate  
$\gamma  $  $/ $  $T_{c }^{0} $ , for the parameters
$\lambda  $ = 1, $\lambda ^{\prime} $ = 0 ,  
k' = 0.5 and for k taking the values  0.01 , 0.1 , 0.2 , 0.5 , 
1.5 and 5  ( k = 1.5 gives exactly an AG law).}
\label{Fig1}
\end{figure}
\begin{figure}
\caption{Experimental results for the variation of the critical  
temperature of YBCO, due to Zn impurities, as a function of  
the residual resistivity. The filled circles are the results  
of Ref.17, the open squares, the crosses, the filled diamonds  
and the filled triangles are respectively the results for the  
batches 2, 3, 4 and 6 of Ref.18 .}
\label{Fig2}
\end{figure}


\begin{references}
\bibitem{41}The symmetric case  $\lambda  $ = $\lambda ^{\prime} $ and 
k = k'  has $x_{2} $ = $x_{1} $  
= $\lambda  $  - k, and is solved explicitely into two branches $x = x_{0} $  
 + $K(\rho ) $ and $x = x_{2} $ with the physical solution corresponding  
to the lower value for x. This degenerate situation is interesting  
to keep in mind, since the general case can be considered as  
resulting from the coupling of these two branches when one moves  
away from this particular case.
\end{references}
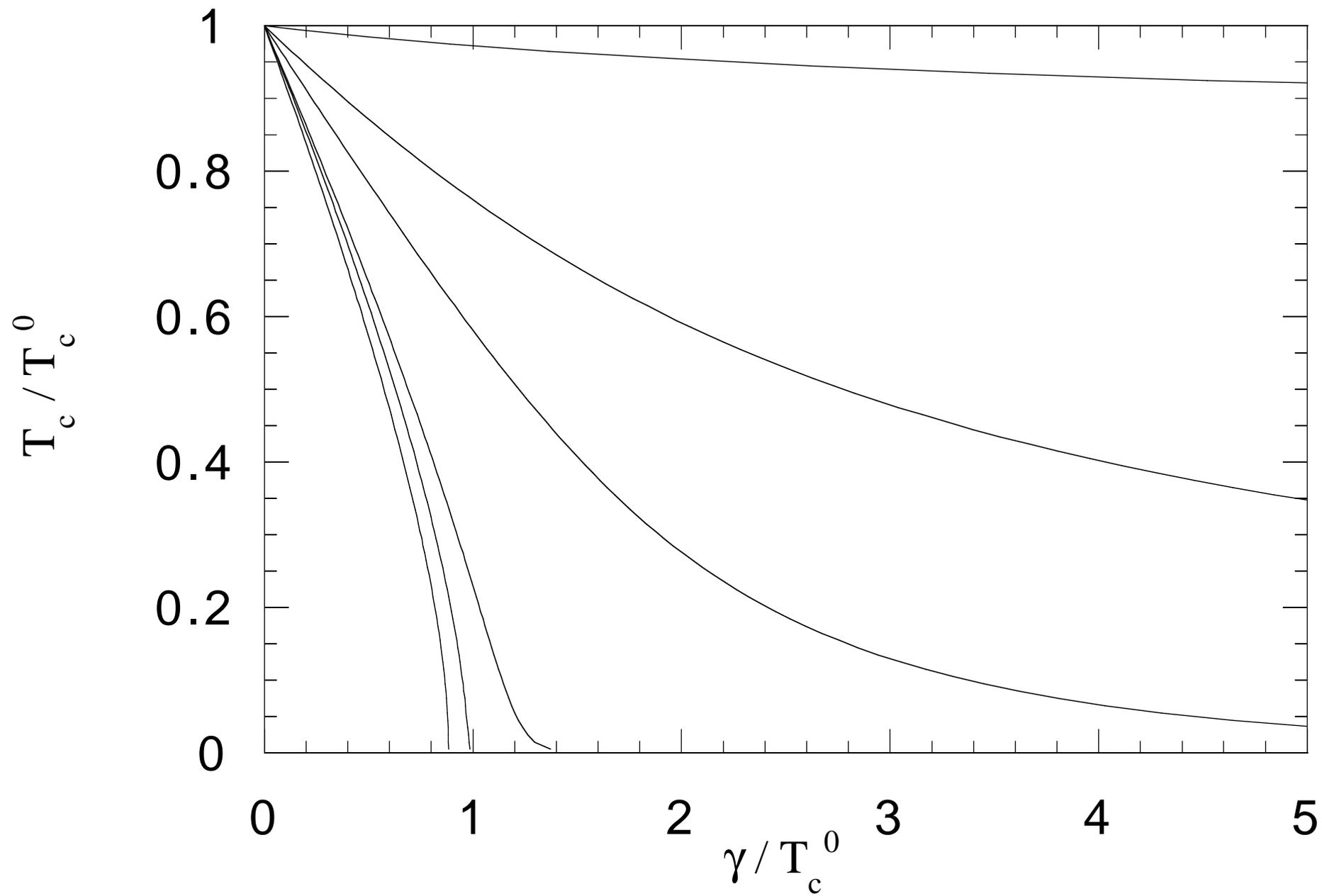
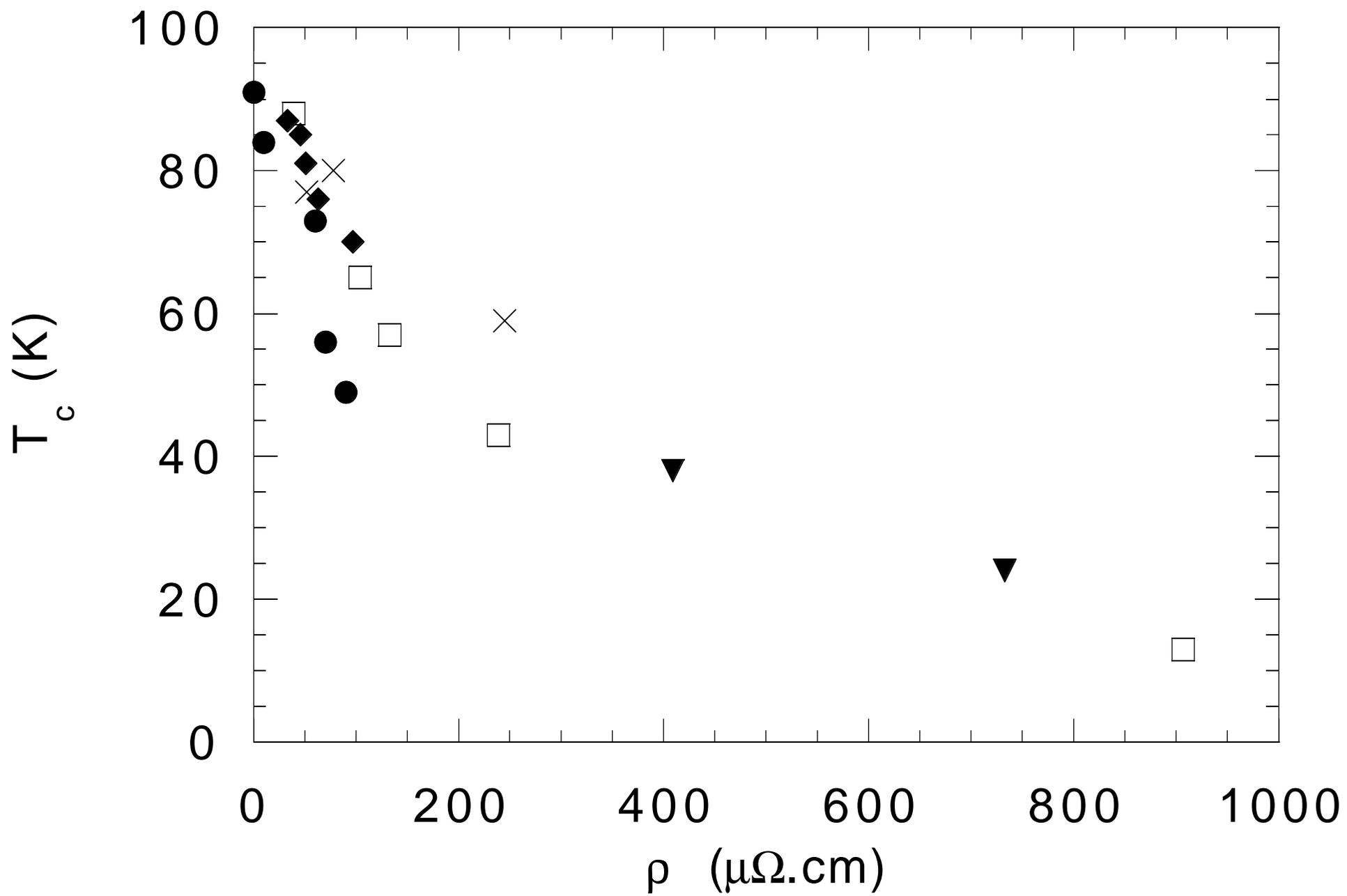
\end{document}